\documentclass[letterpaper]{article} 
\usepackage[submission]{aaai24}  
\usepackage{times}  
\usepackage{helvet}  
\usepackage{courier}  
\usepackage[hyphens]{url}  
\usepackage{graphicx} 
\usepackage{natbib}  
\usepackage{caption} 
\usepackage{algorithm}
\usepackage{algorithmic}
\usepackage{booktabs}
\usepackage{mathtools}
\usepackage{newfloat}
\usepackage{listings}
\DeclareCaptionStyle{ruled}{labelfont=normalfont,labelsep=colon,strut=off} 
\floatstyle{ruled}
\newfloat{listing}{tb}{lst}{}
\floatname{listing}{Listing}
\title{NS4AR: a Focused on Sampling Region Negative\\ Sampling Method in GNN Recommender System}
\author{
    Xiangqi Wang\textsuperscript{\rm 1},
    Dilinuer Aishan\textsuperscript{\rm 2},
    Qi Liu \textsuperscript{\rm 3}
}
\affiliations{

    \textsuperscript{\rm 1 2 3}University of Science and Technology of China\\
    wangxiangqi@mail.ustc.edu.cn, dr517@mail.ustc.edu.cn, 
    qiliuql@ustc.edu.cn
}

\usepackage{bibentry}

\begin{document}

\maketitle

\begin{abstract}
The effectiveness of graphical recommender system depends on the quantity and quality of positive and negative training sets, and especially negative sets. Previous works mainly focused on distributional-based negative sampling method, thus ignoring the topological feature of graphs input, which would be beneficial to integrate information into user-item node if being considered. This paper selects some typical GNN-based recommender system models, as well as some latest negative sampling strategies on the models as baseline. Based on typical GNN recommender system, we divide sampling region into assigned-n areas and use AdaSim to give different weight to these areas to form negative set. Because of the volume and significance of negative items, we also proposed a subset selection model to narrow down the core negative samples.
\end{abstract}

\section{Introduction}

Recommender System can be roughly separated as sampling-based recommender system and non-sampling based recommender system.GNN-based recommender systems are vital in sampling recommender systems because they enable efficient and effective recommendation generation by leveraging the power of graph-based models. By considering the relationships and interactions between users and items, these systems can provide more accurate and personalized recommendations. GNN-based recommender system is vital in various widely-applied sampling-based recommender systems, like Pinsage\cite{Wu2020SelfsupervisedGL}, LightGCN\cite{He2020LightGCNSA} and NGCF\cite{Wang2019NeuralGC}. 

Yet a key challenge remains in graph-based recommendations where only positive pairs are observed in the user-item graph, while other unconnected items are considered to be unobserved negative pairs. Seriously, the number of globally unobserved goods is often large, and it is impractical to count all unobserved negative pairs.

Negative sampling has been widely used in previous work, and the sampling strategy only involves picking a small subset of commodities from a globally unobserved area of goods as negative samples and training the model to distinguish between positive and negative samples, which is proved efficient by\cite{Zhang2013OptimizingTC},\cite{Yuan2021OnEO}. And commonly, a classical strategy is to use a uniform distribution for negative sampling, and the core challenge of such task is sample hard negative samples. However, these graph-based recommendations focus only on the design of the negative sampling distribution and ignore the selection of the sampling region in the GNN information propagation on mechanism. The intuition can be described as GNNs aim to learn node representations that capture the underlying graph structure. By using sampling region-based negative sampling, the GNN model can learn more informative representations. The negative samples from the sampling region help in training the model to distinguish between positive and negative nodes, enhancing the discriminative power of the learned representations.

In this paper, we proposed a novel way of negative sampling based on its graphs' topological region, with subset selection method and flexible sampling-region separating method in addition. This method can separate all data equally into assigned-n regions. From the separated regions, we give them different weights to the layers and nodes and form positive set and negative set.

Since as \cite{Chen2020ASF} has proved the core problem of negative sampling is to learn useful
representations and finds that just the hardest 5\% negatives
are both necessary and sufficient for the downstream tasks.Based on the two sets, we additionally use subset-selection method to sample core negative samples.

The effectiveness analysis of assigned-n region weighted sampling compared to baselines will also be given in following sections.The contribution of this paperwork is based on the often neglected topological feature of the graph dataset and narrow down the core negative set.

\section{Framework}

\begin{figure}[h]
    \centering
    \includegraphics[width=\linewidth]{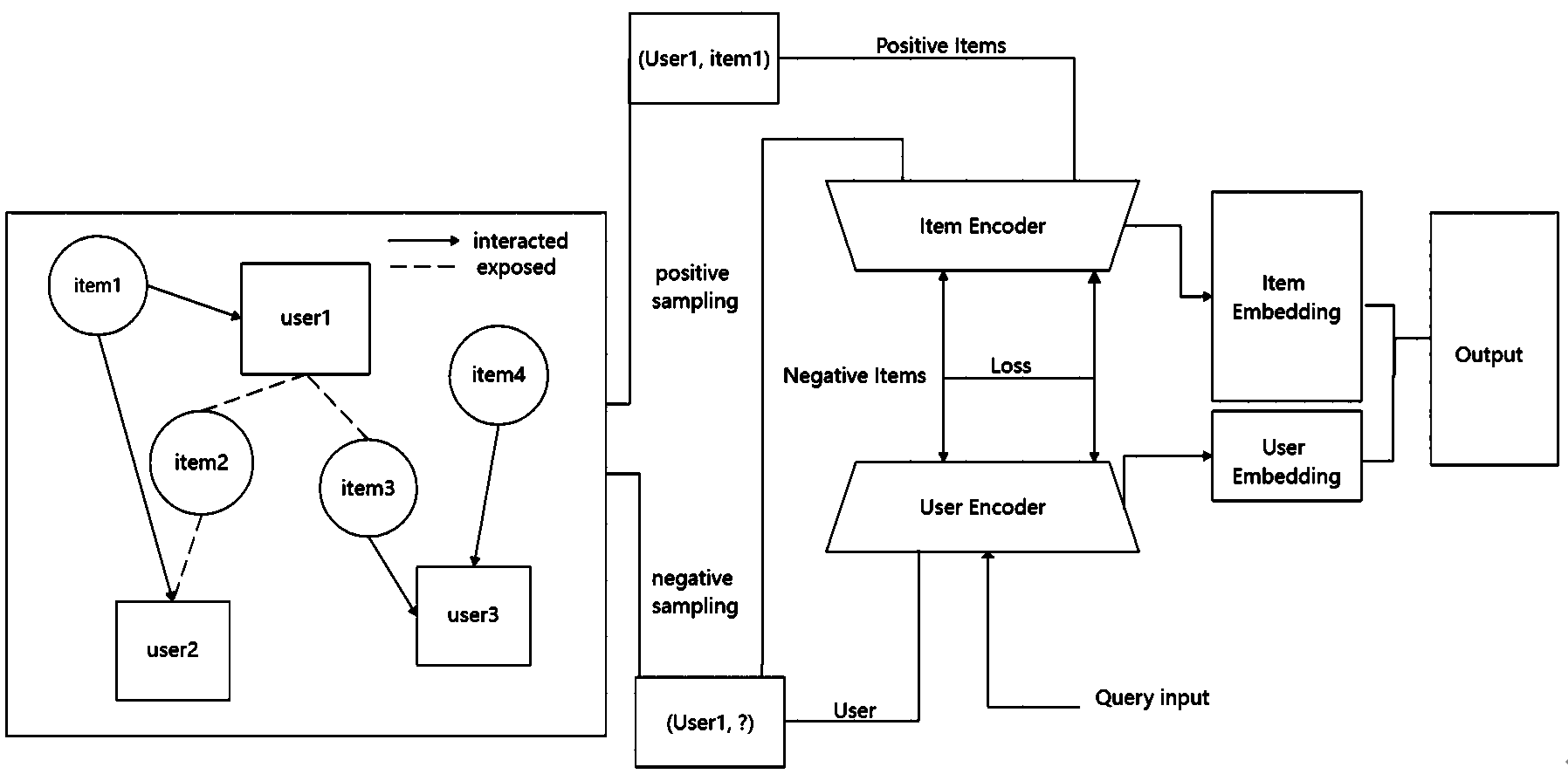}
    \caption{An illustration of general graph-based recommendation, }
    \label{fig:example}
\end{figure}

This algoeithm is based on the typical GNN recommendation pipeline, (see Figure \ref{fig:example}) , which includes a GNN-based $E_{\theta}$ to learn embeddings for items and users. By giving positive sampler and negative sampler in any given user. The sampled user-item interactions serve as the training data for graph-based recommendation learning with stochastic gradient descent (SGD) optimizer. After training, the recommender system recommends the top K items with largest $E_\theta^{u}E_\theta^{v}$ for a queried user.

As show in Figure 1, we gave an example of how GNN recommender system works based on user-item interaction graph and sampled user or item data from it. It also showed how the data got encoded in the encoder, being embedded and used for top-K recommendation.

\subsection{GNN based Encoders}
GNN based encoder can be generally summarized into three main modules.

\textbf{Aggregation Module:}
We maintain an initial item and embedding matrix ${E}_{V} \in {R}^{N \times d}$ and a user embedding matrix ${E}_{U} \in {R}^{M \times d}$. A look-up operation is applied to form an initial embedding vector ${e}_{u} \in {R}^{d}\left({e}_{v} \in {R}^{d}\right)$, where $d$ denotes the embedding dimension. Intuitively, there are two types of aggregation operations: item and user aggregations:

$$
\begin{aligned}
& \mathbf{h}_{u}=\operatorname{Agg}_{u \leftarrow v}\left(\mathbf{e}_{v} \mid v \in \mathcal{S}\left(\mathcal{N}_{u}\right)\right), \\
& \mathbf{h}_{v}=\operatorname{Agg}_{v \leftarrow u}\left(\mathbf{e}_{u} \mid u \in \mathcal{S}\left(\mathcal{N}_{v}\right)\right) .
\end{aligned}
$$

Over here, we use  $\mathcal{N}_{u} / \mathcal{N}_{v}$ denotes neighbors of the central user $u /$ item $v . \mathbf{h}_{u}$ and $\mathbf{h}_{v}$ are the aggregated embeddings for user $u$ and item $v$ respectively. $\mathrm{Agg}_{u \leftarrow v} / \mathrm{Agg}_{v \leftarrow u}$ is the user/item aggregation function. $\mathcal{S}(\cdot)$ represents neighbor sampler.

\textbf{Propagation Module}
To capture higher-order interactions between user and item, we stack multiple propagation layers to propagate embeddings layer by layer. Let $\mathbf{h}_{u}^{l} / \mathbf{h}_{v}^{l}$ represents user/item embedding at the $l$-th layer. The embeddings in $(l+1)$-th layer depends on neighbor's embeddings at $l$-th layer and its own embedding at $l$-th layer. Mathematically, the user embeddings at $(l+1)$-th layer $\mathbf{h}_{u}^{l+1}$ can be defined as:

$$
\begin{aligned}
& \mathbf{h}^{l+1}=\operatorname{Agg}_{u \leftarrow v}\left(\mathbf{e}_{v}^{l} \mid v \in \mathcal{S}\left(\mathcal{N}_{u}\right)\right), \\
& \mathbf{h}_{u}^{l+1}=f\left(\mathbf{h}^{l+1}, \mathbf{h}_{u}^{l}\right) .
\end{aligned}
$$

where $f(\cdot)$ is an update function. Similarity, the item embedding vector at $(l+1)$-th layer also be represented by the abovementioned propagation module.

\textbf{Prediction Module}
After propagating with $L$ layers, we obtain every layer representations as followings: ${\{\mathbf{h}_{u}^{1}, \cdots, \mathbf{h}_{u}^{L}\}} /\left\{\mathbf{h}_{v}^{1}, \cdots, \mathbf{h}_{v}^{L}\right\}$ for the user $u /$ item $v$, respectively.
We utilize the representations of all layers to obtain the final user/item embeddings $\mathbf{e}_{u}^{*} / \mathbf{e}_{v}^{*}$ for prediction can be formulated as:
$$
\begin{aligned}
& \mathbf{e}_{u}^{*}=g\left(\mathbf{h}_{u}^{1}, \cdots, \mathbf{h}_{u}^{L}\right), \\
& \mathbf{e}_{v}^{*}=g\left(\mathbf{h}_{v}^{1}, \cdots, \mathbf{
h}_{v}^{L}\right) .
\end{aligned}
$$

where $g(\cdot)$ denotes a fusion function.

Finally, we use the common way, inner product, to estimate the user's preference towards the target item:

$$
\hat{r}_{u v}=\mathbf{e}_{u}^{*} \cdot \mathbf{e}_{v}^{*}
$$

\section{Sampling method}
Different from the traditional sampling strategy like BPR\cite{Rendle2009BPRBP} and hard negative samplers, like GAN-based samplers\cite{Mirza2014ConditionalGA}. The previous graph-based negative sampling method paid more attention to the design of the negative sampling distribution and ignored the sampled area. 

For example FastGCN\cite{Chen2018FastGCNFL} suggests sampling neighbors in each convolutional layer and AS-GCN\cite{Huang2018AdaptiveST} suggests an adaptive layer-wise neighbor sampling approach.Previous work RecNS in TKDE'22 proposed three-region principle of sampling in sampled area. \\In this paper, we generalize it to N-region principle in sampled area and based on it, since our assign-ed N can be relatively high, by tuning the N as parameter we can guarantee a better performance. The novel subset selection and N-regions inner connection and exploration will be showed below.

Additionally, to explain for the reasons for interpret sampling, we  use social influence theory why these small hops of neighbors can improve the performance.\cite{friedkin_1998},\cite{10.1145/1401890.1401897}.

\subsection{Analysis of negative sampling}

In this subsection,we analyze negative sampling from it
erative GNNs and variance perspectives. The idea of iterative GNNs is to propagate information
layer by layer in the user-item graph to generate user/item
embeddings. 

Yet this might cause over-smoothing problem, take LightGCN \cite{He2020LightGCNSA} as example, its experimental results
demonstrate that the performance begins to decrease after
reaching the peak point on layer 2 in most cases when the
layer number increases from 1 to 4. Actually, the over-smoothing is
inevitable when deepening the network layers but we can
design a more effective negative sampling method to further
improve recommendation performance. . In this paper, we
utilize graph structure to sample negative items in terms of
structural similarities. In the user-item graph, the neighbors
in smaller hop have higher chances of being related to the
central node, which can improve performance by propagating information in smaller-hop neighbors. It shows that the
information propagated in smaller-hop neighbors is more
likely to be positive than negative for the central node. How
ever, propagating information in the higher-hop neighbors
results in performance degradation, which illustrates that
these neighbors information is harmful to recommendation
performance.

Given on the above reasons, negative items should be sampled from some specific region rather than the global unobserved region. From the perspective of negative
sampling, the smaller hop of neighbors (from one-hop to
three-hop) indicates a positive preference, improving the
performance of graph-based recommendations. Thus, the region for negative sampling should be
separated by the propagation mechanism in iterative GNNs.

\subsection{Baseline: 		Three-region principle}
    \begin{figure}[h]
      \centering
      \includegraphics[width=\linewidth]{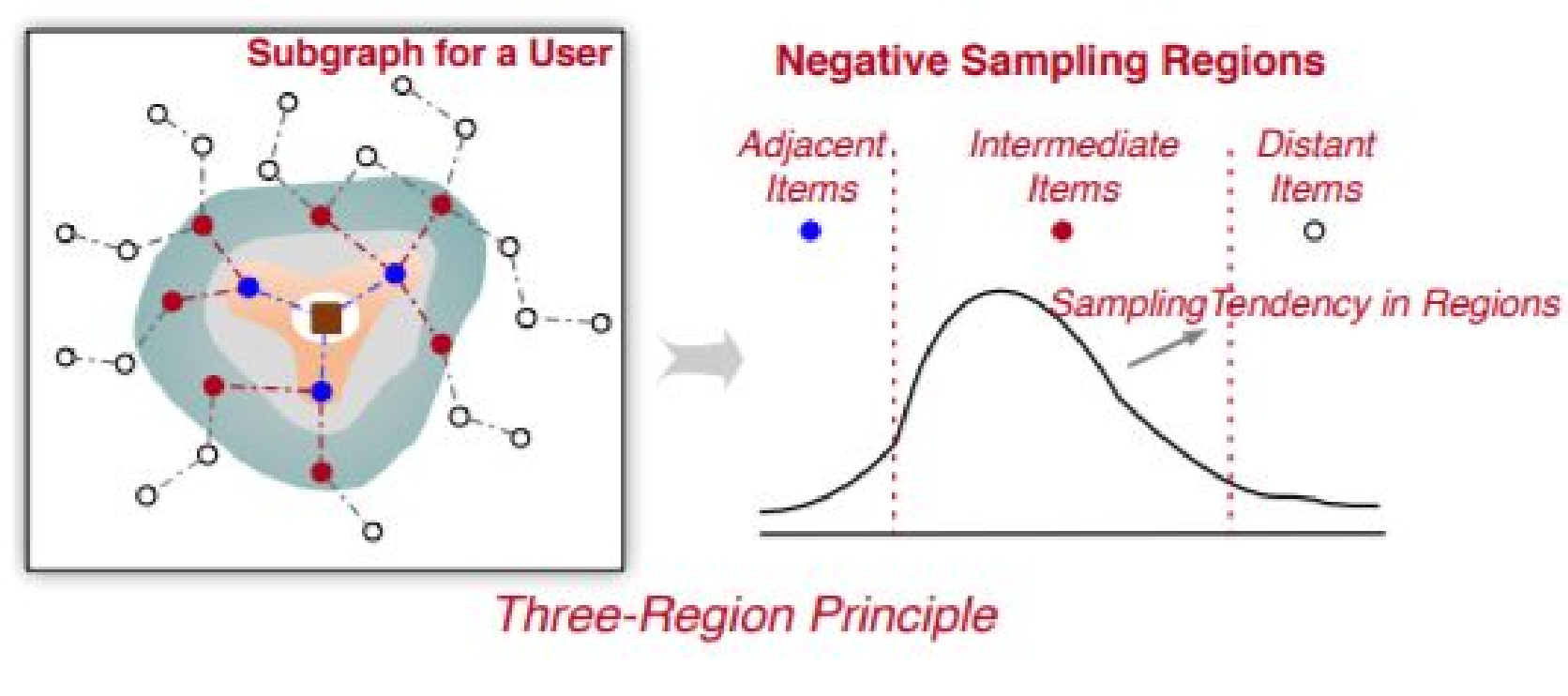}
      \caption{Three region proposed in RecNS}
      \label{fig:example1}
    \end{figure}
As shown in Fig. 2., separating the data into three separate regions, so-called as adjacent items, intermediate items and distant items. Sampling Adjacent points are determined as positive bias of users, and distant method as strong negative bias. Intermediate region is the core region. These intermediate nodes are a bit far from the central user u, and when they propagate information through the user-item graph as adjacent goods, they degrade (or slightly improve) recommendation performance.\\
In this method, sampling more in intermediate regions and less in distant regions is the given in the baseline, yet to what accurate portion in sampling still needs to be determined. These intermediate products offer limited performance gains and sometimes even degradation compared to adjacent products. In addition, propagating these middlewords to the central node can result in significant memory consumption. Therefore, intermediate items should be adequately sampled as negative samples to enhance negative samples.

\subsection{Novel method:      N-region principle}
Different from the three region principle. In out method, we designed a automatic method to traverse the region into $n$ parts and designed a way to select the best $n$. $n$ is a number strictly set, and it's the core parameter to be tuned in the training process.

Intuitively, we can say that the value of 
$n$ determines the granularity or level of detail in which the region is divided. A smaller value of 
$n$ would result in larger and more generalized parts, while a larger value of 
$n$ would result in smaller and more specific parts.

Once the optimal value of 
$n$ is determined, our automatic method traverses the region and divides it into 
$n$ parts accordingly. Each part is then analyzed and processed individually, taking into account the specific characteristics and context within that part.

\begin{figure}[h]
      \centering
      \includegraphics[width=\linewidth]{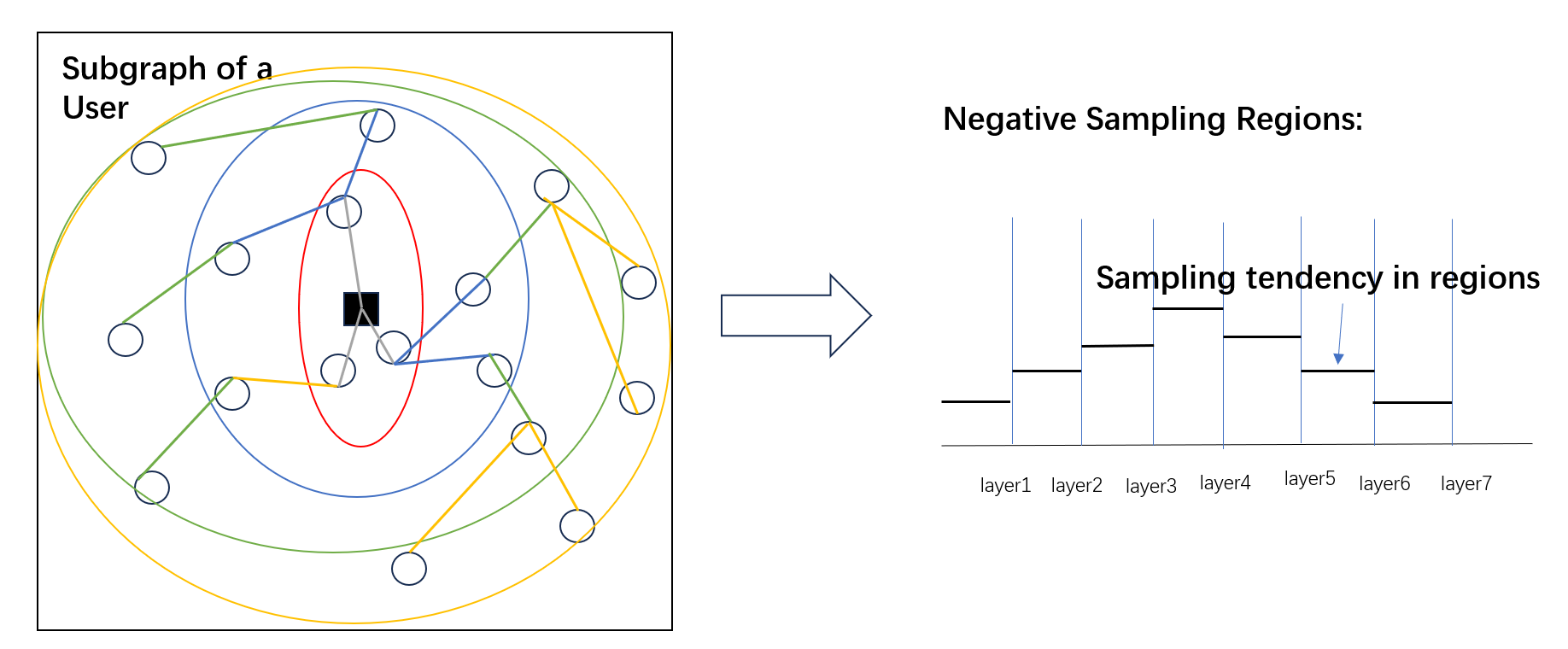}
      \caption{N-regions method sketch}
    \end{figure}

\subsection{Proof of viability of N-region sampling method}

Given parameters are shown in table 1.
Proof is as followings:
Take all positive and negative samples as $Set_{positive}=\Sigma_{i=0}^{n}{||w_{ij}||^{2}_{i}}\times{u^{*}}$\\
$Set_{negative}=\Sigma_{i=0}^{n}{(1-||w_{ij}||^{2}_{i})}\times{u^{*}}$\\
To compare it to baseline method, the exposed yet not clicked as core negative items method in RecNS. Based on it we can prove N-sampling method's superiority\\
Denote $M_u$ as exposed not clicked set and $\beta$ as summing score, which don't count if in intermediate area. The judgement ratio is beta function of exposed data.\cite{Amatriain2015DataMM}
selection method is
$v_n^e=argmax_{v_i\in M_u}\sigma(\beta(e^{*}_u,e^{*}_v))$\\
In traditional method $\beta $ is 1 if $v_e \notin C_u$ and would be the num of exposed items if $v_e \in C_u$.\\
We improve the information validity to tranform $C_u$ into $Set_{negative}$. As we suppose the core negative features of graph as P, then we can give that $AdaSim(C_u,P) \leq AdaSim(Set_{negative},P)$, and in the degree of similarity, $Set_{negative}$ is better than $C_u$.
\begin{table}
  \caption{Table of parameters}
  \label{tab:freq}
  \begin{tabular}{cc}
    \toprule
    Parameter name&Comments\\
    \midrule
    G(v,E) & Graph with v,E as nodes,edges\\
    u&User node u\\
    n &number of regions \\
    $w_{i,j}$ & weight of item i and j connection\\
    $E_v, E_u$& item and user embedding\\
    $h_u^i, h_v^j$&each layer of user(u) or item (v)\\
  \bottomrule
\end{tabular}
\end{table}

\subsection{Essential of additional subset selection on N-region sampling}

In most GNN recommender systems, the volume of negative training embedding is more vital and has more number than positive training embedding. Yet, in most GNN-based recommender system, we require the negative volume to be not too greater than positive set, otherwise there may occur degradation in the final output.   

Having a balanced ratio between the positive and negative samples allows the model to learn from both types of examples effectively. It ensures that the model captures the important patterns and relationships present in the positive samples while also considering the diversity and variety of negative samples.

Therefore, in GNN-based recommender systems, it is essential to carefully control the volume of negative training embeddings, ensuring that it is not excessively larger than the positive set. This balance helps to maintain the overall performance and effectiveness of the recommender system.

\section{Algorithm implemention}
We can describe the algorithm for calculating similarity using N-region sampling and obtaining the set of negative samples $Set_{negative}$, followed by the subset selection process to refine $Set_{negative}$.
\subsection{Optimization function}
We use g( · ) to represent a fusion function, then let $e^{*}_u=g(h^1_u, ... , h^L_u)$, and let  $e^{*}_v=g(h^1_v, ... , h^L_v)$. We use inner product to represent user's preference of item, $r_{uv}=e^{*}_u\times e^{*}_v$.  We choose the hinge loss to optimize the parameters of the GNNs-based encoders, for each user, we sample k negative items, k is $Num_{negative}/Num_{user}$. Then eventual loss function would be:

$L=\frac{1}{k}\times \Sigma_{(u,v)\in O^{+} and v_n \in  S_n(u,v)}[\sigma(\Sigma_{i=1}^{k} e^{*}_u \times e^{*}_{v^{i}_n})-\sigma(k\times e^{*}_u \times e^{*}_v)+ \gamma ]_{+}$

Here, $O^{+}$ represents the set of positive items, Sn represents the NS4AR sampler, $k$ is the number of negative items sampled per user, $e^{*}_{u}$ and $e^{*}_{v}$ are the embeddings of the user and item respectively, and $\gamma$ is the pre-defined hinge-loss margin.

The optimization function aims to maximize the margin between the scores of positive and negative samples. It encourages positive samples to have higher scores than negative samples by at least the margin $\gamma$. The sigmoid function $\sigma$ is used to map the scores to a probability-like value between 0 and 1.

The optimization process involves iteratively updating the parameters of the GNN-based encoders using stochastic gradient descent (SGD) optimizer. The loss function is computed for each batch of training data, and the gradients are backpropagated to update the model parameters. This process continues until convergence or a predefined number of epochs.

By optimizing the parameters of the GNN-based encoders, NS4AR aims to learn better embeddings for users and items, which in turn leads to improved recommendation performance.

\subsection{Basic algorithm implemention}
First shows the algorithm of separating the region into given n different parts. We use drill-by-layer breadth-first search (LBFS\cite{Vigneshwaran2020ClusterBM}\cite{Beamer2012DirectionoptimizingBS}) to traverse subgraphs and get personalized product sets. Here, we define goods within U's K-order neighbors as adjacent goods, and by ascending order to put every U's K-order neighbors in an array.
\begin{algorithm}[!ht]
    \renewcommand{\algorithmicrequire}{\textbf{Input:}}
	\renewcommand{\algorithmicensure}{\textbf{Output:}}
	\caption{BLFS}
	\label
{blfs}
	\begin{algorithmic}[1]
	    \REQUIRE  The User-Item Graph $G(U+V,O^{+})$,khop
	    \ENSURE   The queue array of nearest top nk region $Q_{res}$
	    \FORALL{user u}
	        \STATE {$hop\_num=0,queue=[u]$}
	        \WHILE {$hop\_num<khops$}
	            \STATE {y=BFS{G,queue}}
	            \STATE {queue=y}
	            \STATE {$hop\_num+=1$}
	            \IF{$ hop\_num==khop $}
	                \STATE{$Q_{res}$.append(G-BLFS(y,khop))}
	            \ENDIF
	       \ENDWHILE
	   \ENDFOR
    \end{algorithmic}
\end{algorithm}

\begin{algorithm}[!ht]
    \renewcommand{\algorithmicrequire}{\textbf{Input:}}
	\renewcommand{\algorithmicensure}{\textbf{Output:}}
	\caption{The Breadth-First Search(BFS)}
	\label{blfs}
	\begin{algorithmic}[1]
	    \REQUIRE  The User-Item Graph $G(U+V,O^{+})$,queue
	    \ENSURE   The later-wise node set y
	    \FOR{each node in queue}
	        \STATE{temp=queue.pop(0)}
	        \STATE{nodes=G[temp]}
	        \STATE{Add nodes to y}
	    \ENDFOR
    \end{algorithmic}
\end{algorithm}

\subsection{NS4AR and its training process}
\subsubsection{subset-selection}
In terms of quantity, the exposure of unclicked sample data and difficult negative sample data is much larger than the positive sample, this method iteratively propagates the embedding in the user-commodity graph according to the intrinsic relationship between the samples, so these negative sample strategies are merged into the embedding space. We used several statistical subset-selection method and compared their results in the following charts in Experiments part.\\
We use $f$ to denote subset-selection function, and the final result would be as followings. Suppose u as the m*k hop nodes.\\
$u^{*}=f(Q_{res},u,m,w_i)$\\
Take Backward Stagewise Selection\cite{Evans1986StagewiseDA} as example subset selection method.\\
Take Bayesian information Criterion (BIC)\cite{szmts2022BayesianIC} as judgement rate.\\
Then f would be $argmax_{k=1,2,...p}fisher(Q_{res},u,m,w_i)$\\
The model we use is based on fisher accuracy examination\cite{Connelly2016FishersET}. The fisher check is used to test whether the results of a random experiment support a hypothesis for a random experiment.\\ The details are as follows: if the probability of a random event occurring is less than 0.05, the event is considered to be a small probability event. The general principle is that under a certain hypothesis, the results of a random experiment will not appear with small probability events. If there is a small probability event as a result of a randomized experiment, the hypothesis is deemed not supported.\\
As we use the stagewise model-selection method. Algorithm runs as the followings.\\
\begin{algorithm}[!ht]
    \renewcommand{\algorithmicrequire}{\textbf{Input:}}
	\renewcommand{\algorithmicensure}{\textbf{Output:}}
	\caption{Subset selection method}
	\label{blfs}
	\begin{algorithmic}[1]
	    \REQUIRE  $Q_{res}$,u,m,Weight matrix 
	    \ENSURE   The selected set of $u^{*}$
        \FOR{residual is larger than given threshold}
	       \STATE The coefficient for each feature initially is 0 and the initial  residual is y.
            \STATE stage-wisely given a subset selection option $E$ and use fisher check to check 
            \STATE Each time select the variable with the greatest BIC score with y from all the features, then move the selected variable in a small position in the direction of the variable's regression, and then generate a new residual assigned to y
        \ENDFOR
    \end{algorithmic}
\end{algorithm}
\subsection{$weight_{ij}$ calculation and assisted sampling}
The calculation of $w_{ij}$ is the main contribution of the paper. NS4AR calculate weight on the n-th region which reflects the relevance of the $i^{th}$ adjacent nodes and the $j^{th}$ adjacent nodes. To sum it up, it is kind of way to calculate nodes similarity in complex network.\cite{Hamedani2021AdaSimAR}\cite{Jiang2020NodeSM}\\
We take function $\Phi$ that $\Phi(x)=y$, and y is the label or data of the exact purchasing or useful label.\\
Let $rate_{ij}=\Sigma_{u\in {N_{i}\cap N_{j}} }\frac{1}{(ln(N(u))}$ \\
let $ratio_{ij}=\Sigma_{u\in {N_{\Phi{i}}\cap N_{\Phi{j}}} }\frac{1}{(ln(N(\Phi{u}))}$\\
let final result $weight_{ij}=rate_{ij}\times{log(ratio_{ij})}+ratio_{ij}\times{log(rate_{ij})}$\\
This is equivalent to increasing the weight on the calculation of common neighbors,\cite{Wang2022CommonNM}\cite{Ahmad2020MissingLP} lowering the weight if Common neighbor has more neighbors, otherwise increasing the weight.\\
\begin{algorithm}[!ht]
    \renewcommand{\algorithmicrequire}{\textbf{Input:}}
	\renewcommand{\algorithmicensure}{\textbf{Output:}}
	\caption{Training of NS4AR(exposed not clicked recommendation)}
	\label{blfs}
	\begin{algorithmic}[1]
	    \REQUIRE  The queue array of nearest top nk region $Q_{res}$, subset selection function $f$, 
	    \ENSURE   The accurate recommend point $v_e$,
	    \FOR {i=1,2,...n}
	        \FOR {j=1,2,....n}
	            \STATE calculate $weight_{ij}=rate_{ij}\times{log(ratio_{ij})}+ratio_{ij}\times{log(rate_{ij})}$ 
	        \ENDFOR
	    \ENDFOR
	    \FOR{i=1,2,...n}
	        \STATE use specified subset selection method,$u^{*}=f(Q_{res},u,m,w_i)$, to change nodes cluster $u$ into $u^{*}$
	    \ENDFOR
	    \STATE use weight matrix to formulate $Set_{negative}$ and $Set_{positive}$
	    \STATE create $C_u$ use given set
	    \STATE $v_n^e=argmax_{v_i\in C_u}\sigma(\beta(e^{*}_u,e^{*}_v))$
	    \STATE return $v_n^e$
    \end{algorithmic}
\end{algorithm}

\subsection{Time Complexity:}
The computation complexity of NS4AR comes from these parts. In the exposure yet not clicked NS4AR recommendation process, (1)the computational complexity of
the self-amplified factor $\beta$. The time for this part is $O(M)$.(2) The inner AdaSim calculation is $O(Md)$.(3) The subset selection process is $O(Mlogd)$.(4)The time cost for sampling one exposed negative item
is $O(Md)$. Thus the total complexity result would be $O(Md)$

\section{Experiments}
\subsection{Experimental Settings}
\textbf{Datasets:} We conducted the method on two datasets: Zhihu and Alibaba datasets. For each user u, we randomly select 80\%
of user’s interactions as the training set and use the next 10%
of interactions as the validation set for hyperparameters
tuning and early stopping. The remaining 10\% interactions
are used as the test set to evaluate the performance. \\
Zhihu is a large QA website, where users click on in-
terested articles to read. Here, we use a public dataset
released in CCIR-2018 Challenge1, which contains both
article exposure information and user click information.\\
Alibaba dataset collects user behaviors from the E-
commerce platform Taobao. We sample a subset of data
that contains users’ historical interactions, including posi-
tive interactions and exposure information.\\
KHOP is the assigned n number, is defaultly assigned as 100. In the experiment, khop is approximately 100 is the best n to assign.
\subsection{Baselines and Frameworks}
We use LightGCN and Pinsage as framework, take some negative sampling strategies as baselines. Negative Sampling strategies include: RecNS, SRNS\cite{Kipf2016SemiSupervisedCW}, MCNS\cite{Yang2020UnderstandingNS}, SimNS\cite{Ding2019ReinforcedNS}, DNS\cite{Rendle2014ImprovingPL}. Accurate sampling strategies description can be found in the according cited papers. parameter settings would be as follows: The embedding size is fixed to 64 for LightGCN, while PinSage is set to 256. We implemented based on Tensorflow 2.0 and Python 3.10. The extra parameters remain as default.\\
\subsection{Performance Comparision}
We summarize the detailed performance comparison
among all negative sampling strategies on the Zhihu and
Alibaba datasets in Table 2. \\
Exact difference can be easily summarized.
Changing khop(assigned-n) and the result would be as followings in Table.3 \\
\begin{table*}[t]
\centering
\caption{Result of NS4AR and some several negative sampling strategies(from top to down represent the result of MCNS,SimNS,DNS,RecNS,NS4AR}
\label{table1}
\begin{tabular}{|c|c|c|c|c|c|c|c|c|c|c|c|}
\hline
\multicolumn{3}{|c|}{LightGCN Zhihu}&\multicolumn{3}{c|}{PinSage Zhihu}&\multicolumn{3}{c|}{LightGCN Alibaba}&\multicolumn{3}{c|}{PinSage Alibaba}\\
\hline
Recall&NDCG&HR &Recall&NDCG&HR &Recall&NDCG&HR&Recall&NDCG&HR \\
\hline
4.36&5.10&52.66&2.42&2.72&34.55&5.44&2.51&7.92&3.02&1.34&4.58\\
\hline
4.42&5.12&53.16&2.22&2.62&35.75&5.24&2.81&8.02&3.22&1.54&4.88\\
\hline
4.40&5.08&53.25&2.31&2.47&35.66&5.34&2.80&8.11&3.11&1.61&4.91\\
\hline
4.47&5.13&53.19&2.29&2.39&35.64&5.29&2.77&8.13&3.15&1.65&4.95\\
\hline
4.59&5.77&54.27&2.36&2.49&35.75&5.33&2.82&8.22&3.24&1.72&5.16\\
\hline
\end{tabular}
\label{table_MAP}
\end{table*}
\begin{table*}[t]
\centering
\caption{Changing NS4AR assigned-n from 1 to 1000, sequence is 1, 10, 100, 1000}
\label{table1}
\begin{tabular}{|c|c|c|c|c|c|c|c|c|c|c|c|}
\hline
\multicolumn{3}{|c|}{LightGCN Zhihu}&\multicolumn{3}{c|}{PinSage Zhihu}&\multicolumn{3}{c|}{LightGCN Alibaba}&\multicolumn{3}{c|}{PinSage Alibaba}\\
\hline
Recall&NDCG&HR &Recall&NDCG&HR &Recall&NDCG&HR&Recall&NDCG&HR \\
\hline
3.25&4.22&45.24&1.99&2.41&30.25&5.01&2.42&7.72&2.82&1.14&4.18\\
\hline
4.23&5.00&53.17&2.29&2.45&35.33&5.19&2.61&7.79&3.02&1.44&4.52\\
\hline
4.46&5.15&53.26&2.31&2.43&35.66&5.33&2.80&8.18&3.22&1.62&5.06\\
\hline
4.32&5.03&53.11&2.37&2.21&35.51&5.20&2.75&8.11&3.19&1.55&4.88\\
\hline
\end{tabular}
\label{table_MAP}
\end{table*}
\subsection{Comparsion analysis and furthur analysis}
NS4AR can greatly improve HR and Recall values, but it does not have much impact on NDCG\cite{Chia2021BeyondNB}, perhaps because it has used positive example auxiliary sampling\cite{Tsai2022ConditionalCL} in the RecNS process, and the already positive example has considerable weight in the overall sampling result ratio.
To compare the result furthur, we also included the data of training and training charts in following.\\
\subsubsection{Does N-region principle improve negative sampling}
To verify the impact of our method, we designed an experiments that samples the assigned-n candidates from all regions. The experimental results will be shown in Fig.3.\\
Take n as 5, this chart represents the the Recall@20 result on Pinsage and LightGCN. Only sampling examples in distant regions contributes little to the final result, which should be avoided. The nearest positive k nodes also contribute little. Taking 4th and 5th together would improve a lot.
\begin{figure}[h]
      \centering
      \includegraphics[width=0.75\linewidth]{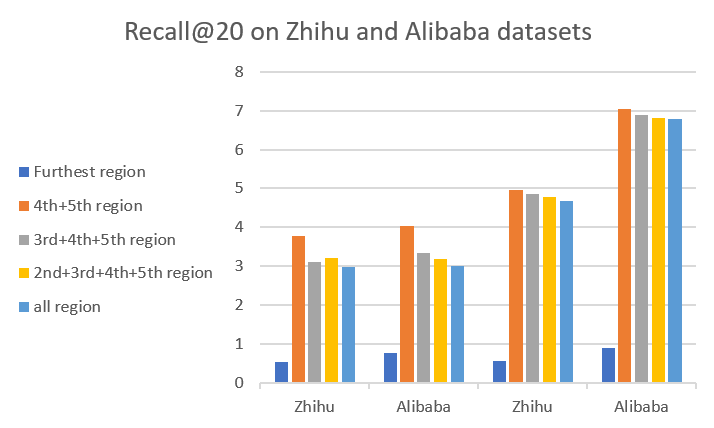}
    \end{figure}
\subsubsection{How could possibly change the network for furthur boost in performance}
Sampling negative samples based on the sampling-area is an interesting target to go. More and more strategies focused on sampling areas or taking advantages of graphical topological features should be investigated. To furthur exploit the real graphical based recommendation dataset, fully using of the exposured yet not interacted data would be of great help. Due to the multiple queries and real applications, change the model to online learning or reinforcement learning would be an promosing topic. 

\section{Conclusion}

The paper proposes a novel method called NS4AR (Focused on Sampling Region Negative Sampling Method in GNN Recommender System) for negative sampling in GNN-based recommender systems. The key contributions of the paper are as follows:

1.\textbf{N-region principle}: The paper introduces the concept of dividing the sampling region into N regions and assigning different weights to these regions. This allows for more fine-grained sampling and improves the quality of negative samples.

2. \textbf{Subset selection}: The paper proposes a subset selection model to narrow down the core negative samples. This helps in reducing the volume and significance of negative items, leading to better performance.

3.\textbf{Weight calculation and assisted sampling}: The paper presents a method to calculate the weight of item connections based on the similarity between nodes. This weight calculation helps in determining the relevance of adjacent nodes and guides the sampling process.

The experimental results on Zhihu and Alibaba datasets demonstrate that NS4AR outperforms several baseline negative sampling strategies in terms of recall, NDCG, and HR. The paper also discusses the impact of different values of N (assigned-n) on the performance.

Potential future work could include exploring other sampling strategies based on the sampling region, further optimizing the subset selection process, and investigating online learning or reinforcement learning approaches for GNN-based recommender systems. Additionally, the paper suggests fully utilizing the exposed but not interacted data and incorporating more graphical topological features for improved performance.

\bibliography{aaai24}

\end{document}